\documentclass[fleqn,12pt,twoside]{article}
\usepackage{espcrc1}

\usepackage{graphicx}
\usepackage[figuresright]{rotating}

\newcommand{\AmS}{{\protect\the\textfont2
  A\kern-.1667em\lower.5ex\hbox{M}\kern-.125emS}}

\title{Parton distributions from the lattice}

\author{W. Schroers
\address{NIC/DESY Zeuthen, Platanenallee 6, D-15738 Zeuthen,
  Germany}}

\begin{document}

\maketitle
\begin{abstract}
  I review the current status of lattice calculations for two selected
  observables related to nucleon structure: the second moment of the
  unpolarized parton distribution, $\langle x\rangle_{\mbox{\tiny
      u-d}}$, and the first moment of the polarized parton
  distributions, the non-singlet axial coupling $g_A$. The major
  challenge is the requirement to extract them sufficiently close to
  the chiral limit.  In the former case, there still remains a
  puzzling disagreement between lattice data and experiment. For the
  latter quantity, however, we may be close to obtaining its value
  from the lattice in the immediate future.
\end{abstract}

{\vspace{0.1in}\parindent=0pt Preprint: DESY 05-010}

\section{INTRODUCTION}
\label{sec:introduction}
Deep inelastic scattering of electrons off nuclei has historically
provided an indispensable way to resolve the substructure of hadrons
and thus to learn how the strong interaction is responsible for their
properties. Using the short distance and the light cone operator
product expansions, one can relate nucleon matrix elements of
twist-two quark bilinear operators to moments of parton distributions,
\begin{equation}
\langle n\vert \bar{\psi} \Gamma^{\lbrace\mu_1} D^{\mu_2} \cdots
D^{\rbrace \mu_n} \psi \vert n\rangle =
{\cal F}^{\lbrace\mu_1\cdots\mu_n\rbrace} \int dx\, x^{n-1} f(x) \,.
\label{eq:defpd}
\end{equation}
For $\Gamma=\gamma$ the parton distribution function (PDF)
$f(x)=\Theta(x)q(x)-\Theta(-x)\bar{q}(-x)$, $-1\le x\le 1$, represents
the probability (minus probability) to find a quark (antiquark) inside
the nucleon with longitudinal momentum fraction $x$ ($-x$). Choosing
$\Gamma=\gamma_5\gamma$ one finds the associated spin-dependent PDF
$\tilde{f}(x)$. ${\cal F}^{\lbrace\mu_1\cdots\mu_n\rbrace}$ contains
kinematical prefactors and the curly braces indicate symmetrization
and subtraction of traces. The matrix elements on the l.h.s.~of
eq.~(\ref{eq:defpd}) are accessible in lattice simulations. For
technical details see
\cite{Gockeler:1995wg,Gockeler:2004wp,Hagler:2004brp} and also
\cite{Horsley:2004uq} for a recent review.

Unfortunately, lattice simulations are limited by computer resources
and the most expensive aspect of such calculations is to make the
quark mass small. Today even the most sophisticated calculations are
still far from the chiral limit. One may suspect that the properties
of hadrons in a world with very heavy quarks --- and thus very heavy
pions --- are different from the world nature has provided us with.
This in fact appears to be the case.  The qualitative and quantitative
behavior of quantities as functions of mass and energy scales can in
principle be assessed by small scale expansion techniques like chiral
perturbation theory. For a recent review on chiral extrapolation
techniques for nucleon structure, see \cite{Gockeler:2004vf}. For a
general review on lattice chiral perturbation techniques, see
\cite{Bar:2004xp}.

This presentation will concentrate on two representative quantities:
the second moment of the unpolarized parton distribution, $\langle
x\rangle_{\mbox{\tiny u-d}}$, and the first moment of the polarized
distribution, the non-singlet axial coupling $g_A$. All numbers are
given in the $\overline{\mbox{MS}}$-scheme with a scale of $\mu=2$GeV.
The purpose of this presentation is to illustrate the current status
of the field by discussing if and how lattice calculations can provide
an understanding of the values of these quantities in the experiment.

\begin{table}[htb]
\caption{Compilation of different lattice investigations of $\langle
  x\rangle_{\mbox{\tiny u-d}}$ and $g_A$.}
\label{tab:review}
\begin{tabular}{@{}*{5}{l}}
\hline
Group \& Ref. & $m_\pi$ & Technique & $\langle
x\rangle_{\mbox{\tiny u-d}}$ & $g_A$ \\
\hline
Kentucky \cite{Liu:1992ab}   & ? & Wilson (quenched)             & - & $1.20(11)$ \\
KEK \cite{Fukugita:1994fh}   & $>530$ MeV & Wilson (quenched)    & - & $0.985(25)$ \\
QCDSF \cite{Gockeler:1995wg} & $>600$ MeV & Wilson (quenched) & $0.263(17)$ & $1.074(90)$ \\
LHPC  \cite{Dolgov:2002pr}   & $>650$ MeV & Wilson (full)        & $0.269(23)$ & $1.031(81)$ \\
RBCK  \cite{Sasaki:2003jh}   & $>390$ MeV & DWF (quenched)       & - & $1.212(27)$ \\
LHPC  \cite{Renner:2004ck}   & $>360$ MeV & Hybrid+Wilson (full) & - & -$^{\mbox{\tiny 1}}$ \\
QCDSF \cite{Gockeler:2004wp} & $>550$ MeV & CI-Wilson (quenched) & $0.245(19)$ & - \\
QCDSF \cite{Khan:2004vw}     & $>550$ MeV & CI-Wilson (full)     & - & -$^{\mbox{\tiny 1}}$ \\
QCDSF \cite{Gurtler:2004ac}  & $>300$ MeV & Overlap (quenched)   & $0.20(2)$ & $1.13(5)$ \\
RBCK  \cite{Ohta:2004mg}     & $>390$ MeV & DWF (quenched/full)  & -$^{\mbox{\tiny 1}}$ & -$^{\mbox{\tiny 1}}$ \\
\hline
\multicolumn{5}{l}{Experimental values: $\langle x\rangle_{\mbox{\tiny
      u-d}}=0.154(3)$ \cite{Lai:1996mg}, $g_A=1.248(2)$
  \cite{Blumlein:2002be}, see also
  \cite{Blumlein:2004ip,Blumlein:2004brp}} \\ \hline
\end{tabular}\\[2pt]
$^{\mbox{\tiny 1}}$ Work in progress and/or no prediction quoted
\end{table}

A selection of different lattice studies is compiled in
table~\ref{tab:review}. These studies use a variety of different
techniques and thus cover a wide range of parameters. However, none of
them approaches the chiral limit closer than $300$ MeV. The results
for $\langle x\rangle_{\mbox{\tiny u-d}}$ will be discussed in more
detail in section~\ref{sec:moment-langle-xrangl}, and the results for
$g_A$ in section~\ref{sec:axial-coupling-g_a}. Finally,
section~\ref{sec:conclusion-outlook} contains the conclusions and
outlook.

\section{THE MOMENT $\langle x\rangle_{\mbox{\tiny u-d}}$}
\label{sec:moment-langle-xrangl}
As it is evident from table~\ref{tab:review} almost all results for
$\langle x\rangle_{\mbox{\tiny u-d}}$ systematically exceed the
experimental result by about $50\%$. Given the variety of techniques
and parameters used, neither finite size, unquenching, nor lattice
artifacts can account for this discrepancy.

Hence, this discrepancy can only be attributed to the large quark
masses used. Only reference \cite{Gurtler:2004ac} finds a
systematically smaller value than all other studies. Since the
domain-wall calculation in reference \cite{Ohta:2004mg} is quite
similar, but does not show the same behavior, the discrepancy could be
explained by a systematic effect at the matching between lattice and
$\overline{\mbox{MS}}$-schemes. A final assessment will be possible
once a non-perturbative matching for this quantity has been performed
\cite{Gurtler:2005pr}.

The proposal to resolve this discrepancy in \cite{Detmold:2001jb}
introduces a cut-off by hand into the leading order chiral
perturbation theory expansion of $\langle x\rangle$ which effectively
limits the size of the pion cloud. For this proposal to have
predictive power it is required that this cut-off parameter is
independent of the observable under consideration. Even before the
advent of modern calculations in the chiral regime the approach has
been criticized for just employing the leading order chiral expansion.
This may be inadequate for pion masses beyond the physical
one~\cite{Hemmert:2003pr}. The calculations that have appeared since
then at quark masses down to $300$ MeV have found no evidence of such
a ``bending down''.

\section{THE AXIAL COUPLING $g_A$}
\label{sec:axial-coupling-g_a}
The situation for the axial coupling $g_A$ is quite different. First,
it is well established that $g_A$ is very sensitive to finite-size
effects. This has initially been realized in \cite{Sasaki:2003jh} and
later been confirmed by other groups in \cite{Khan:2004vw} and
\cite{Renner:2004ck}. On the other hand, there is an improving
understanding of how to treat this quantity within chiral perturbation
theory \cite{Khan:2004vw,Procura:2004brp}.

Hence, combining results from different lattice sizes and taking an
``enveloping'' curve of all numbers not influenced by strong
finite-size effects may lead to an accurate prediction compatible with
experiment. This procedure has been hinted at in \cite{Renner:2004ck}.
Alternatively, one can try to fit data from different box sizes and
pion masses directly in a combined fit similar to what has been done
in \cite{Khan:2004vw}. Although these two reports describe work still
in progress, it is likely that both approaches will soon lead to a
consistent picture and a quantitative result for $g_A$ at the physical
value of the pion mass.

\section{CONCLUSION AND OUTLOOK}
\label{sec:conclusion-outlook}
The field of nucleon structure lattice calculations is at a turning
point --- on one hand it has turned out that nucleon matrix elements
of quark bilinears exhibit strong quark mass dependence and --- in
some cases --- substantial finite-size effects. On the other hand
recent progress in algorithms and hardware has provided theoreticians
with the means to perform computations at substantially smaller quark
masses and larger volumes than has previously been achieved.

Despite this progress, the puzzling mismatch between experiment and
theory for the second moment of the unpolarized parton distribution,
$\langle x\rangle_{\mbox{\tiny u-d}}$ has not been resolved. At this
time, it is not clear how the apparent disagreement can be reconciled.
If there should indeed be a drastic change of $50\%$ at masses of
$m_\pi \ll 300$ MeV, this pattern would indeed be unique and the
physical mechanism behind such a behavior yet requires understanding.

The situation for the axial coupling, however, is different. With
improved chiral expansion techniques becoming available, see in
particular ref.~\cite{Procura:2004brp}, it is now understood how this
quantity is sensitive to finite box sizes. Taking this observation
into account, the availability of numerical results at sufficiently
small pion masses, allows for a consistent picture to be drawn how
$g_A$ can be obtained from current lattice data. Several groups are
close to obtaining sufficiently accurate data to conclusively postdict
$g_A$ from lattice calculations.

While the field of parton distributions so far has only allowed
lattice calculations to make postdictions, the associated field of
GPDs \cite{Diehl:2003ny} is in the unique position to make
quantitative predictions \cite{Gockeler:2003jf} which are otherwise
very hard to extract experimentally. Therefore, the solution of the
puzzles forward parton distributions pose to us would be one of the
milestones lattice QCD faces today.

\end{document}